\documentclass[preprint2]{aastex}
\newcommand{\vi}{\hbox{$V\!-\!I$}}
\newcommand{\uv}{\hbox{$U\!-\!V$}}

\newcommand{\vk}{\hbox{$V\!-\!K$}}

\newcommand{\ubv}{\hbox{$U\!BV$}}

\begin{document}

\title{Testing population synthesis models with globular cluster colors}

\author{Pauline Barmby \& John P. Huchra}
\affil{Harvard-Smithsonian Center for Astrophysics, 60 Garden St., Cambridge, MA 02138} 

\shorttitle{Models and cluster colors}
\shortauthors{Barmby \& Huchra}

\begin{abstract}

We have measured an extensive set of $UBVRIJHK$ colors for M31 globular
clusters \citep{b00}. We compare the predicted simple stellar population colors of three 
population synthesis models to the intrinsic colors of Galactic and 
M31 globular clusters.
The best-fitting models fit the cluster colors very well -- the
weighted mean color offsets are all $<0.05$ mag.
The most significant offsets between model and data are in the $U$ and 
$B$ passbands; these are not unexpected and are likely due to
problems with the spectral libraries used by the models.
The metal-rich clusters (${\rm [Fe/H]}\gtrsim-0.8$) are best fit by 
young (8 Gyr) models, while the metal-poor clusters 
are best fit by older (12--16 Gyr) models.
If this range of globular cluster ages is correct, it
implies that conditions for cluster formation must have existed 
for a substantial fraction of the galaxies' lifetimes.

\end{abstract}


\section{Introduction}

Predicting the integrated spectral energy distributions of stellar populations
is important in the solution of many problems in astronomy, from 
determining the ages of globular clusters to modeling counts of faint galaxies
at high redshift. Beginning with the early work of \citet{tin68}, 
successive generations of modelers have combined the best available
data on stellar structure and evolution to predict
the appearance of the combined light of generations of stars.
Although the subject of population synthesis has a long history,
it is an active area of research: synthesis techniques and many
of the input data (isochrones, opacities, spectral libraries)
continue to be improved.

There is good evidence that globular clusters (GCs) are internally homogeneous in age 
and metallicity \citep{h99,ste93}. GCs are the best observational analogs
of modelers' `simple stellar populations', i.e.\ populations of stars formed
over a short time out of gas with homogeneous chemical composition.
Broadband colors are among the simplest predictions of population
synthesis models, so comparing the models' predicted colors
to cluster colors is the natural zeroth-order test of compatibility between
the models and reality \citep{huc96}. 
In this paper we compare the broad-band $UBVRIJHK$ colors predicted by 
three modern population synthesis models with the colors of Galactic
and M31 GCs. Our observational database is the first one
to include extensive coverage of the $JHK$ bandpasses, and the first
with spectroscopic metallicities for all clusters. 
We use the cluster metallicities to bin the clusters
for comparison to the appropriate models.
In this way we determine the cluster-to-model offset 
separately for each color and avoid the ambiguity 
in comparing model and cluster colors in two-color diagrams.

\section{Input data and comparison procedure}

For M31 clusters, the observational data are from the \citet{b00}
catalog. For the Galactic clusters we obtained 
optical colors, metallicity, and reddening from
the June 1999 version of the \citet{h96} catalog, and IR colors
from \citet{fpc80}, as reported in \citet{bh90} (but with
the reddening correction applied by \citet{bh90} removed 
and the reddening values in \citet{h96} used instead).
We dereddened the clusters' colors
using the values of $E_{B-V}$ given in the catalogs and the
\citet{ccm89} extinction curve for $R_V=3.1$. 
For M31, we excluded clusters where the error in the spectroscopic 
metallicity was ${\sigma}_{\rm [Fe/H]}>0.5$, and clusters
suspected of being young on the basis of strong Balmer absorption or
blue \bv\ colors \citep[see][]{b00}.
For both galaxies, we excluded clusters with $E_{B-V}>0.5$; 
there are 103 M31 and 85 Galactic clusters in the final sample.
Photometric data is not available in all bandpasses for all clusters:
only about two-thirds have measured $R$ and $I$, and less than
half have $H$.

\begin{figure}
\includegraphics*[scale=0.9, angle=0]{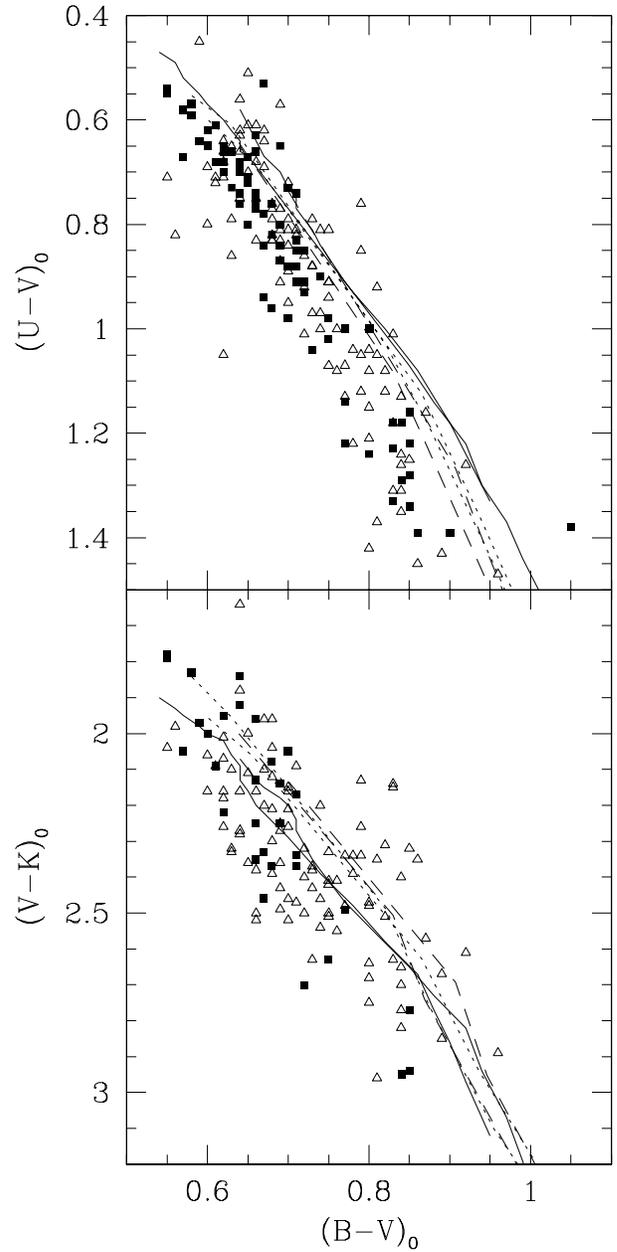}
\caption{$(B-V)_0$ vs. $(U-V)_0$ and $(V-K)_0$ for M31 globular clusters (triangles) and Galactic GCs (squares).
Lines are population synthesis models of ages 8~Gyr (bluer colors) and 16~Gyr (redder colors): 
Worthey (solid), BC (dashed), KFF (dotted).\label{twocolor}}
\end{figure}

We compare the cluster colors to those for simple stellar 
populations of ages 8, 12, and 16 Gyr from three sets of models:
those of Worthey\footnote{The version we used updates the \citet{w94} 
models by including
a more realistic treatment of the horizontal branch for ${\rm [Fe/H]}<-1.0$.}, 
Bruzual and Charlot (hereafter BC) \citep[both the Worthey and BC models are
reported in][]{lei96}, and \citet{kff99} (hereafter KFF).
Although model colors are tabulated in smaller age increments
(typically 1 Gyr), initially it is more reasonable to use the models
as a rough guide to relative ages rather than attempting to derive precise
cluster ages from them.
The Worthey models are computed at [Fe/H] values of
$-2.0$,$-1.5$,$-1.0$,$-0.5$, and $-0.25$ dex, and the BC and
KFF models are computed at [Fe/H] values of $-2.33$ (KFF models only), 
$-1.63$, $-0.63$, and $-0.32$ dex.
We compared clusters to
both the Salpeter IMF (Worthey's `vanilla' models) and \citet{sca86}
\citep[in the Worthey models]{ms79} IMF version of the models.
\citet{w94} finds that some of his model colors have defects 
(e.g.\ \bv\ is too red by 0.04-0.06$^m$ due to problems in the
theoretical stellar atmospheres and the color-temperature calibration),
but the sizes of these defects are not well-determined so
we do not correct for them. Figure~\ref{twocolor} shows data
and models in two frequently-used two-color diagrams.

Since the models are computed at discrete values of [Fe/H], we use
the spectroscopic metallicities of the clusters to 
compare only clusters with comparable metallicities
($\pm0.25$~dex) to each model.
The Galactic cluster metallicities given in \citet{h96} are
on the \citet{zw84} (ZW) metallicity scale, and the M31 cluster metallicities
are also tied to this scale through the calibration of \citet{bh90}.
Recent work \citep{cg97,rhs97} suggests that the ZW scale may be
non-linear at both high and low metallicities. 
We retain the ZW scale in this paper because we found 
that using the \citet{cg97} scale to assign clusters
to model comparison bins made little difference in our results.
We caution, however, that the effect of changing the metallicity scale 
is unknown for the ${\rm [Fe/H]}=-0.25$ model bin. 
The transformation from the ZW to CG scales 
is only defined for ${\rm [Fe/H]_{ZW}}<-0.5$, 
the lower limit of this metallicity bin. 

\begin{figure}
\includegraphics*[scale=0.9,angle=0]{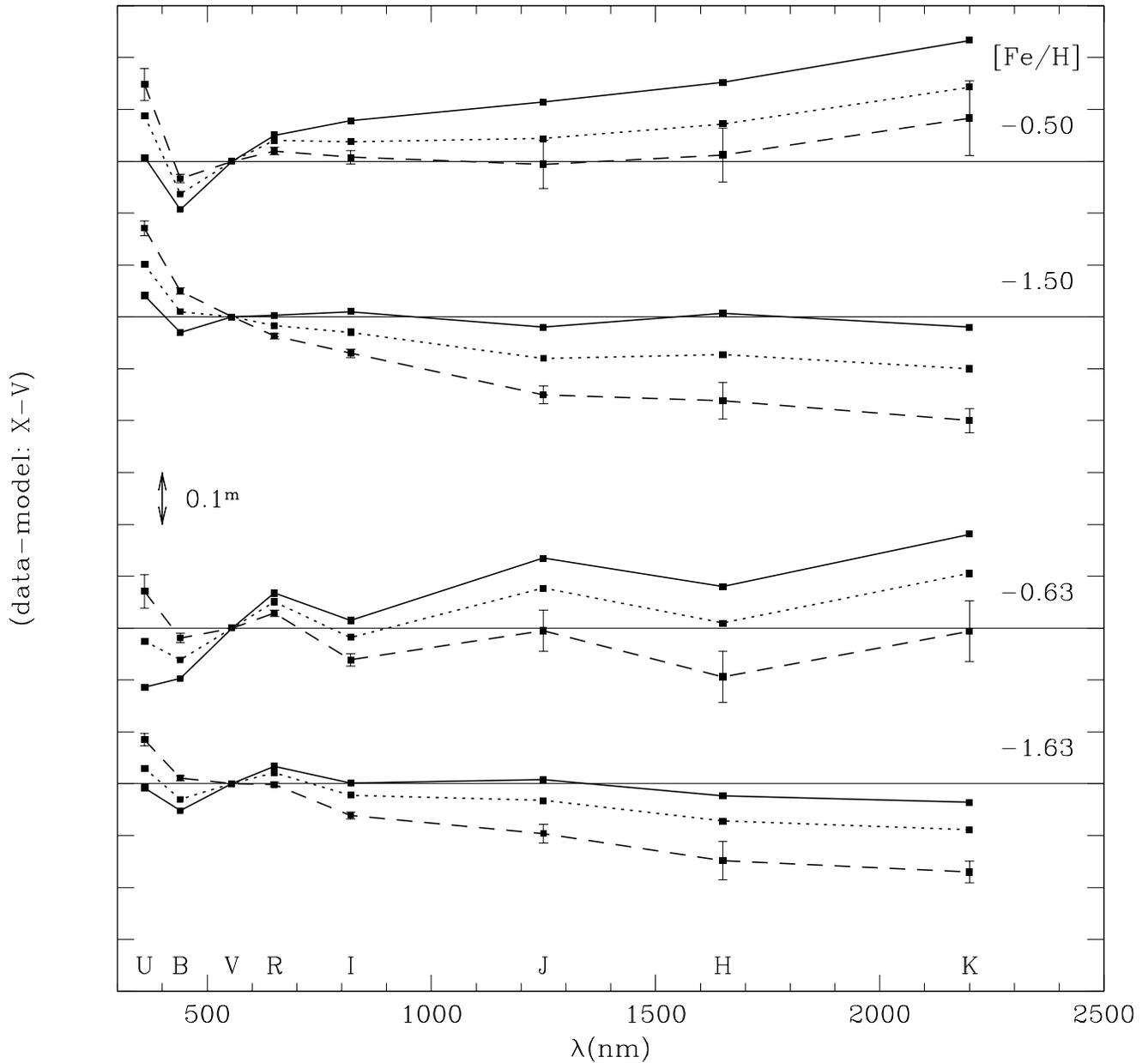}
\caption{Color offsets $\Delta(X-V)$ (data$-$models) for Salpeter IMF. 
($\Delta V=0$ is plotted to
emphasize that the models are normalized to the data at $V$.) 
Solid lines: 16 Gyr models, dotted lines: 12 Gyr models, dashed lines: 8 Gyr models. 
The $-1.5$ and $-0.5$ bins in [Fe/H] are Worthey models; the
$-1.63$ and $-0.63$ [Fe/H] bins are BC models. Error bars (plotted only on the 8 Gyr models for clarity) 
are the standard errors of the mean cluster colors and do not include observational uncertainties. 
\label{salp-zwscale}}
\end{figure}

\begin{figure}
\includegraphics*[scale=0.9,angle=0]{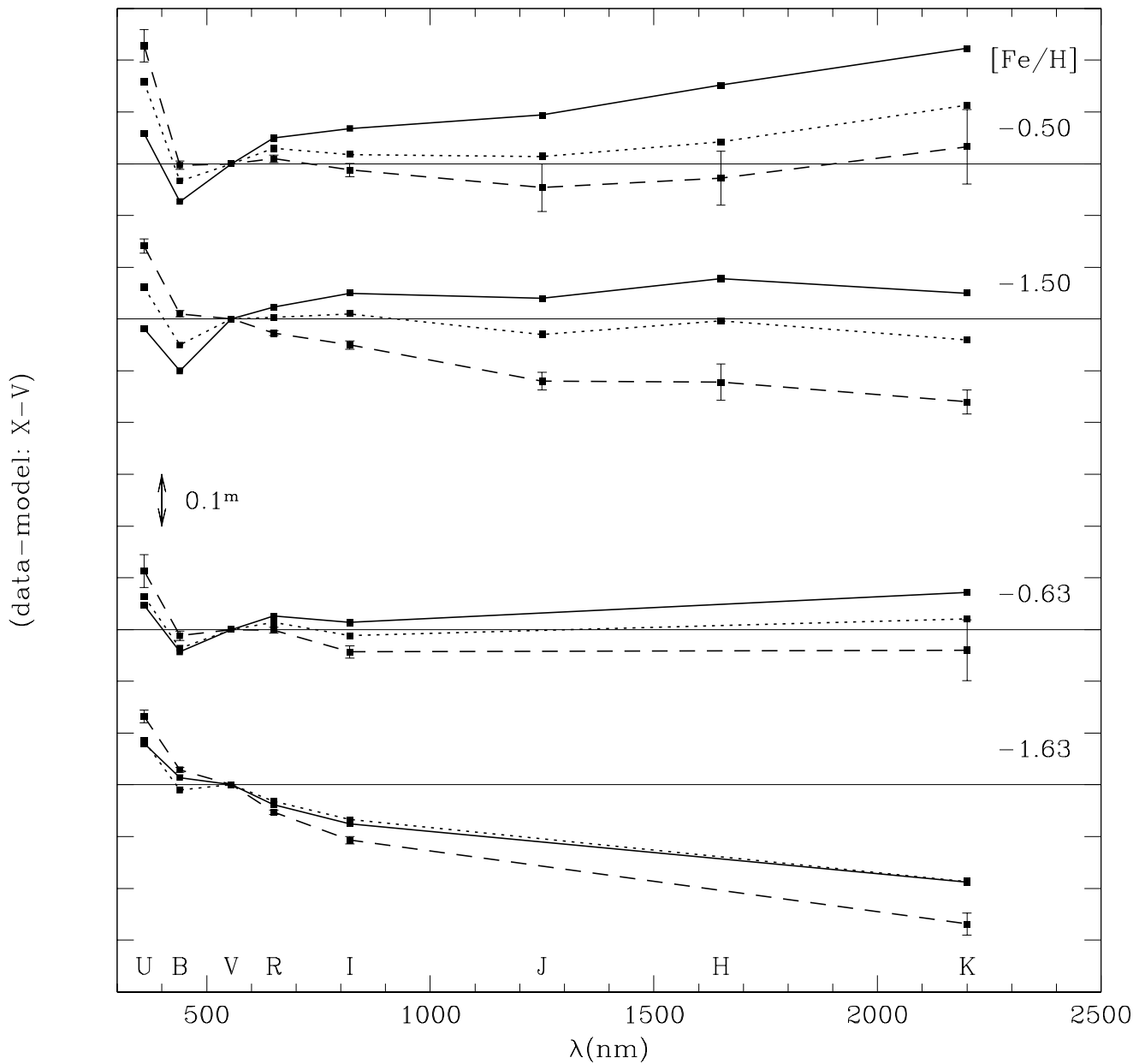}
\caption{Color offsets $\Delta(X-V)$ or $\Delta(V-X)$ (data$-$models)
for Scalo or Miller-Scalo IMF. Symbols
same as Figure~\ref{salp-zwscale}. The $-1.5$ and $-0.5$ bins in [Fe/H] 
are Worthey models; the $-1.63$ and $-0.63$ [Fe/H] bins are KFF models
(which have no $\Delta(J-V),\Delta(H-V)$ since they do not
predict these colors).\label{scal-zwscale}}
\end{figure}

We calculated the mean offsets between model and cluster 
colors (referenced to $V$) 
for each metallicity bin; Figures~\ref{salp-zwscale}--\ref{scal-zwscale} 
show some representative comparisons. 
We plot $\Delta(X-V)$ for all bandpasses $X$ to make clear the 
differences in spectral energy distributions between models and 
data; we remind the reader that the offsets for bandpasses redward of $V$ 
thus have the opposite sign from the usual colors.
One general characteristic of the models visible in the Figures
is that younger-aged models predict bluer colors. The 
exception is the KFF Scalo
model for ${\rm [Fe/H]}=-1.63$, which predicts only very small color 
differences ($\lesssim0.01^m$) between ages of 12 and 16~Gyr.
The effect of the IMF on the
colors appears to depend on both metallicity and age. 
For the Worthey ${\rm[Fe/H]}=-1.50$ models,
Miller-Scalo IMF colors are redder than Salpeter model colors
at all ages,
but for the  ${\rm[Fe/H]}=-0.50$ models, the Miller-Scalo IMF
colors are bluer for 8 and 12~Gyr and almost identical for 16 Gyr.
BC predict almost no color difference between the Salpeter and 
Scalo IMF models of the same age and metallicity.

A striking feature in Figures~\ref{salp-zwscale} 
and~\ref{scal-zwscale} is the range of discrepancies between models
and data. For example, the largest difference between the Worthey model
with parameters (Salpeter IMF, ${\rm[Fe/H]}=-1.50$, age 16~Gyr) and the mean
colors of clusters with $-1.75\leq{\rm[Fe/H]}\leq-1.25$ is $0.04^m$ in \uv. 
The same models with ${\rm[Fe/H]}=-0.50$
are well offset  from the data at all colors except \uv; the largest offset 
is $0.23^m$ in \vk.
To determine the best-fitting models, we quantify the overall goodness-of-fit 
for each model/cluster metallicity bin pair as:
\begin{equation}
F=\frac{\sum_k |\Delta(X-V)_k|/{\sigma}^2_k}{\sum_k 1/{\sigma}^2_k}
\end{equation}
The color differences $\Delta(X-V)_k$ are weighted by 
$1/{\sigma}^2_k$, where ${\sigma}_k$ are the standard errors in
the mean colors of objects in the bin.
Table~\ref{bestfit} gives the $\Delta$ and $F$ values for the 
best fitting models in each metallicity bin.

\section{Discussion}

Table~\ref{bestfit} shows that the best-fitting models fit the
data quite well, with typical color offsets of $0.02-0.03^m$.
The two bandpasses with the most significant offsets are
$U$ and $B$: the models are too blue in \uv\ and too red in \bv. Neither
offset shows a clear trend with metallicity.
The offsets are likely not due to systematics in the
photometric system or in the extinction curve. While problems 
with the photometric systems might be expected in the $R$ and $I$
bands (due to conversion between the Johnson and Cousins 
$RI$ systems), both data and models use the well-defined Johnson \ubv\ system.
Problems in the reddening curve also seem unlikely for the same
reasons. We suspect that the offsets are more likely due to
systematic errors in the models. The \bv\ offset in particular
is likely due to the flux libraries used. 
Both \citet{w94} and \citet{lej97} found their model \bv\ colors 
to be $0.04-0.06^m$ too red compared to empirical solar-metallicity
spectra, even after correcting to the empirical color-temperature scale. 
This suggests a possible problem with the stellar atmosphere models of
\citet{k95}, upon which both libraries are based.

The cause of the offset in \uv\ is not as clear. This offset is actually
worse than it appears: since we compute the \uv\ colors for the BC and KFF
models as $(U\!-\!B)+(B\!-\!V)$, the red \bv\ colors
{\it compensate} for some of the \ub\ defect, which is actually larger
than the defect in \uv. \citet{w94} -- whose models give \uv\ directly --
finds that his model \uv\ is too blue compared to solar neighborhood
stars and elliptical galaxies. 
Worthey cites problems with the $U$ fluxes from the stellar libraries
as a possible cause: modeling the many blended atomic and
molecular lines blueward of $B$ is difficult, and many of the necessary
opacities are not well determined.
This cannot be the only cause of model problems in $U$, since
the BC and KFF models, which use the same stellar library, predict
different \uv\ colors. 
The treatment of the horizontal branch in the models is another possible
source of problems in the \uv\ colors because the HB emits most of the blue light.
However, systematic problems with the model HB color (which depends on metallicity), 
would presumably produce a \uv\ offset dependent on metallicity,
which we do not observe.
Observational error is another possible contributor to the \uv\ offset, as
many of the \ub\ colors of the M31 clusters are poorly determined \citep[see Table 3
of][]{b00}. 
Understanding the rest-frame $U$ flux of stellar populations becomes
increasingly important when studying high-redshift galaxies and
global star formation history, and further investigation of the models
in this bandpass is clearly warranted.

A secondary result in Table~\ref{bestfit} is that
age determines which model best fits the data. Higher-metallicity
cluster colors are best fit by 8 Gyr models, regardless of IMF. 
Lower-metallicity cluster colors (${\rm [Fe/H]_{bin}}\leq-1.00$) are best 
fit by 12 or 16 Gyr models. The best-fit age depends on the IMF
for several of the models, but not in any systematic fashion.
This result is consistent with the determinations of
relative ages for Galactic clusters by \citet{ros99}.
These authors determined relative ages of 35 Galactic globular
clusters from a homogeneous set of $V$, \vi\ color-magnitude
diagrams. They compared theoretical isochrones with the observational
CMDs to determine ages using two independent methods.
They found that the clusters with ${\rm [Fe/H]_{CG}}>-0.9$
were $\sim17$\% younger than clusters with ${\rm [Fe/H]_{CG}}<-1.2$, with
the intermediate-metallicity clusters showing a $\sim25$\% age dispersion.
These results are model-dependent, as are ours, but the 
results' similarity implies that either there is a real 
difference between metal-rich and metal-poor clusters
or there is a systematic problem in the models in
one of the metallicity regimes.

What possible systematic errors in our input data or comparison
procedure could produce the result that the metal-rich clusters
are younger? We redid the comparison procedure considering the
clusters of each galaxy separately, and still found younger ages
for the most metal-rich clusters. Although M31 has a greater 
proportion of the metal-rich clusters, younger ages are
found for both M31 and Galactic metal-rich clusters.
\citet{cm94} suggest that the spectroscopic metallicities 
of the most metal-rich M31 clusters measured by \citet{hbk91}
are systematically too high. If this is true, the clusters would
appear too blue compared to old, higher-metallicity models
and the best-fit model would be younger.
We compared the metal-rich M31 clusters to the Worthey ${\rm [Fe/H]}=-1.0$ models,
and the best-fitting model had age 16~Gyr.
However, the goodness-of-fit was better for the young, metal-rich models 
than for the older, more metal-poor model, so we conclude that
younger ages are still favored for these clusters.
Overestimating the reddening of the metal-rich clusters
would make the derived intrinsic colors too blue and yield younger ages.
This seems unlikely, given that the color-metallicity
relations for Galactic and M31 clusters match well throughout
their metallicity range \citep[see ][]{b00}, and the methods of
reddening determination for M31 and Galactic clusters are different.

If the detection of younger ages for metal-rich globular clusters
is real, it has implications for galaxy formation. A range of GC 
ages implies that GC formation took place over an
extended period of time. Conditions for GC formation 
were not particular to the early universe, an assertion supported
by observations of `proto-globular' clusters in present-day
merging galaxies \citep[e.g.\ ][]{zep99}. 
More precise knowledge of the distribution of cluster
ages in each galaxy would be extremely useful in understanding cluster
system formation. If the age distribution is continuous,
the relation between age and metallicity might hold clues as to
what factors controlled the cluster formation rate.
If the age distribution is bimodal -- with
most clusters old and coeval and the remainder younger and coeval -- 
then some event must have triggered the second episode of
GC formation. Perhaps the younger clusters were
stripped from or accreted along with satellite galaxies of
M31 and the Galaxy.

\section{Conclusions}

Comparison of three sets of population synthesis models with
integrated colors of M31 and Galactic globular clusters shows that the
models reproduce the redder average cluster colors to within the 
observational uncertainties. 
The poorer agreement in \uv\ and \bv\ is likely due to systematic
errors in the models. Younger-age models are required to best
match the colors of the metal-rich clusters, consistent
with the findings of \citet{ros99} that the most metal-rich
Galactic clusters are younger than the bulk of the globular
cluster population. A range of ages for globular clusters implies 
that conditions for cluster formation were not restricted to the 
early universe. The cluster age distribution has important
implications for galaxy and globular cluster system formation,
and attempts to determine it more precisely are needed.

\clearpage
\onecolumn

\clearpage

\begin{deluxetable}{llllrrrrrrr}
\tabletypesize{\footnotesize}
\rotate
\tablewidth{0pt}
\tablecaption{Color offsets for best-fitting models\label{bestfit}}
\tablehead{\colhead{[Fe/H]} & \colhead{model} & \colhead {$F$} & \colhead {$N_{\rm gc}$} & \colhead{$\Delta(U-V)$} & \colhead{$\Delta(B-V)$} & \colhead{$\Delta(V-R)$} & \colhead{$\Delta(V-I)$} & \colhead{$\Delta(V-J)$} & \colhead{$\Delta(V-H)$} & \colhead{$\Delta(V-K)$}}
\tablecolumns{11}
\startdata
-0.25&W8sc & 0.023&15&$ 23.4\pm2.4$&$-0.4\pm0.9$&$-3.0\pm1.1$&$-0.3\pm1.0$&$  3.2\pm 2.6 $&$-1.4\pm 6.0 $&$-6.0\pm4.9 $ \\
-0.32&K8sc & 0.022&17&$ 11.5\pm2.2$&$-2.5\pm0.8$&$-2.1\pm1.0$&$ 0.2\pm0.9$&\nodata&\nodata       &$-3.9\pm4.6 $ \\
-0.50&W8sc & 0.013&30&$ 22.8\pm3.1$&$-0.3\pm0.8$&$-1.0\pm0.7$&$ 1.2\pm1.3$&$  4.6\pm 4.6 $&$ 2.8\pm 5.2 $&$-3.3\pm7.2 $ \\
-0.63&K8sp & 0.012&35&$ 12.2\pm3.2$&$-1.0\pm0.9$&$-0.6\pm0.6$& $2.4\pm1.2$&\nodata&\nodata       &$ 2.1\pm5.9 $ \\
-1.00&W12sc& 0.014&48&$ 13.7\pm2.2$&$-2.3\pm1.0$&$ 0.0\pm0.5$&$-1.1\pm1.2$&$  3.2\pm 1.8 $&$ 6.9\pm 2.5 $&$ 4.3\pm2.1 $ \\
-1.50&W16sp& 0.016&75&$  4.1\pm1.4$&$-3.0\pm0.6$&$-0.3\pm0.5$&$-1.0\pm0.8$&$  2.0\pm 1.7 $&$-0.7\pm 3.5 $&$ 2.0\pm2.3 $ \\
-1.63&K16sp& 0.019&77&$  5.1\pm1.2$&$-0.7\pm0.5$&$ 1.7\pm0.4$&$ 2.8\pm0.7$&\nodata        &\nodata       &$10.8\pm2.1 $ \\ 
-2.00&W16sp& 0.027&37&$  8.2\pm1.2$&$-2.6\pm0.6$&$-0.3\pm0.8$&$-1.6\pm1.2$&$ -6.7\pm 2.6 $&$-5.2\pm 3.7 $&$-5.8\pm3.5 $ \\
-2.23&K16sp& 0.027&12&$  4.1\pm2.3$&$ 1.8\pm1.3$&$ 2.3\pm2.1$& $3.4\pm2.0$&\nodata&\nodata&$-7.1\pm6.2 $ \\
\enddata
\tablecomments{Units are hundredths of a magnitude. The capital letter in the model column indicates the
best-fitting model (Worthey, BC or KFF), the number is the model age in Gyr, and `sp' or 'sc' indicate
Salpeter or Scalo IMF, respectively.}
\end{deluxetable}

\end{document}